\documentclass[prd,twocolumn,nofootinbib,aps,superscriptaddress,tightenlines,preprintnumbers]{revtex4}

\usepackage{epsfig}
\usepackage{amsmath}

  \newcount\hour \newcount\minute
  \hour=\time \divide \hour by 60
  \minute=\time
  \newcommand{\mydate}{\ \today \ - \number\hour :\ifnum \minute<10 0\fi
\number\minute}

\def\nn{\nonumber \\ }

\def\vev#1{\left\langle #1 \right\rangle}

\newcommand{\pbra}[1]{\left(#1\right)}
\newcommand{\bbra}[1]{\left[#1\right]}

\begin{document}

\preprint{\hbox{CALT-68-2747}  }

\title{Inflaton Two-Point Correlation in the Presence of a Cosmic String}

\author{ Chien-Yao Tseng and Mark B. Wise}

\affiliation{California Institute of Technology, Pasadena, CA 91125 }

\begin{abstract}
Precise measurements of the microwave background anisotropy have confirmed the inflationary picture of approximately scale invariant, Gaussian primordial adiabatic density perturbations. However, there are some anomalies that suggest a small violation of rotational and/or translational invariance in the mechanism that generates the primordial density fluctuations.  Motivated by this we study the two point correlation of a massless scalar (the inflaton)  when the stress tensor contains the energy density from an infinitely long straight cosmic string in addition to a cosmological constant.
\end{abstract}
\widetext
\date{\mydate}
\maketitle

\baselineskip=13pt

\section{Introduction}
The inflationary cosmology is the standard paradigm for explaining the horizon problem \cite{Guth:1980zm,{Linde:Albrecht}}. In its simplest form inflation predicts an almost scale invariant spectrum of approximately Gaussian density perturbations \cite{quantumfluct,inflationreview}.  Rotational and translational invariance dictate that the two point correlation of the Fourier transform of the primordial density perturbations $\delta({\bf k})$ has  the form,
\begin{equation}
\label{basic}
\langle \delta({\bf k}) \delta({\bf q})\rangle=P(k)(2 \pi)^3\delta({\bf k}+{\bf q}),
\end{equation}
where $k=|{\bf k}|$ and $P$ is called the power spectrum. In Eq.~(\ref{basic}) the fact that  $P$ only depends on the magnitude of the wave-vector ${\bf k}$ is a consequence of rotational invariance and the delta function of ${\bf k}+{\bf q }$ arises from translational  invariance. Let $\chi ({\bf k})$ be the Fourier transform of a massless scalar field with canonical normalization. Its two point correlation in de-Sitter space is
\begin{equation}
\langle \chi({\bf k}) \chi({\bf q})\rangle=P_{\chi}(k)(2 \pi)^3\delta({\bf k}+{\bf q}),
\end{equation}
where $H$ is the Hubble constant during inflation and
 \begin{equation}
P_{\chi}(k)= { H^2 \over 2k^3}.
\end{equation}
In the inflationary cosmology the almost scale invariant density perturbations that are probed by the microwave background anisotropy and the  large scale structure  of our observed universe have a power spectrum that differs from $ P_{\chi}(k)$ normalization factor that has weak $k$ dependence\footnote{We will treat this factor as a constant and denote it by $\kappa^2$.}  and a transfer function that arises from the growth of fluctuations at late times after they reenter the horizon \cite{{COBE},{BOOMERANG},{ACBAR},{CBI},{VSA},{ARCHEOPS},{DASI},{MAXIMA},{WMAP3}}.

Inflation occurs at an early time when the energy density of the universe is large compared to energy scales that can be probed by laboratory experiments. It is possible that there are paradigm shifts in our understanding of the laws of nature, as radical as the shift from classical physics to quantum physics, that are needed to understand physics at the energy scale associated with the inflationary era. Motivated by the lack of direct probes of physics at the inflationary scale Ackerman {\it et al}  wrote down the general form that Eq.~(\ref{basic}) would take \cite{Ackerman} if rotational invariance was broken by a small amount during the inflationary era  (but not today) by a preferred direction and computed its impact on the microwave background anisotropy (see also \cite{Gumrukcuoglu:2006xj,ArmendarizPicon:2007nr,Pereira:2007yy,Gumrukcuoglu:2007bx,Yokoyama:2008xw,Kanno:2008gn,Watanabe:2009ct}). They also wrote down a simple field theory model that realizes this form for the density perturbations where the preferred direction is associated with spontaneous breaking of rotational invariance by the expectation value of a vector field. This model serves as a nice pedagogical example, however, it cannot be realistic because of instabilities \cite{Himmetoglu}.  Evidence in the WMAP data for the violation of rotational evidence  was found in Ref.~\cite{Pullen:2007tu,Groeneboom:2008fz,ArmendarizPicon:2008yr}.
Another anomaly in the data on the anisotropy of the microwave background data is the ``hemisphere effect" \cite{Eriksen:2003db,{Eriksen:2007pc}}.  This cannot be explained by the model of Ackerman {\it et. al.}. Erickeck {\it et. al} proposed  an explanation based on  the presence of a very long wavelength (superhorizon) perturbation \cite{Erickcek:2008sm}. This long wave-length mode picks out a preferred wave-number and can give rise to a hemisphere effect. It  violates translational invariance  and there are very strong constraints from the observed large scale structure of the universe on this \cite{hirata:ISW,{hirata:lensing},{Hirata:2009ar}}.  The generation of large scale temperature fluctuations in the microwave background temperature by superhorizon perturbations is known as the Grishchuk-Zel'dovich effect \cite{Grishchuk-Zeldovich}.

Carroll, {\it et. al.} proposed explicit forms for violations of  translational invariance \cite{Tseng}, in the energy density perturbations two point correlation,  motivated by: the symmetries that are left unbroken,  the desire to have a prediction for the two point correlation of multipole moments  of the microwave background anisotropy  $\langle a_{lm} a^*_{l'm'}\rangle$ that is  non-zero for at most a few  $l$'s that are different from $l'$, and the desire to introduce at most a few new parameters. To get a feeling for what can happen in general consider a case where there is a special point ${\bf x}_0$  during inflation. Its presence violates translational invariance, however translational  invariance is restored if in addition to translating  the spatial coordinates we also translate ${ \bf x}_0$.  So in coordinate space $\langle \delta( {\bf x})\delta ({\bf y})\rangle$ must be a function of ${\bf x}$, ${\bf y}$ and ${\bf x}_0$ that is invariant under translations  ${\bf x} \rightarrow {\bf x}+{\bf a}$, ${\bf y} \rightarrow {\bf y}+{\bf a}$, ${\bf x}_0 \rightarrow {\bf x}_0+{\bf a}$ and rotations ${\bf x} \rightarrow R{\bf x}$,  ${\bf y} \rightarrow R{\bf y}$, ${\bf x}_0 \rightarrow R{\bf x}_0$. Furthermore it must be symmetric under interchange of ${\bf x}$ and ${\bf y}$.  Ref.~\cite{Tseng} assumed $\langle \delta( {\bf x})\delta ({\bf y})\rangle$ only depends on the two variables,  $({\bf x}-{\bf x}_0)^2+({\bf y}-{\bf x}_0)^2 $and $|{\bf x}-{\bf y}|$, and expanded in the dependence the first of these. However in the general case of a special point ${\bf x}_0$ Eq.~(\ref{basic}) becomes

\begin{equation}
\langle \delta( {\bf k})\delta ({\bf q})\rangle= e^{i({\bf k}+{\bf q})\cdot {\bf x}_0}P(k,q,{\bf k} \cdot {\bf q}),
\end{equation}
where $P$ is symmetric under interchange of ${\bf k}$ and ${\bf q}$. Without further simplifying assumptions about the form of $P$ and the value of ${\bf x}_0$ this will result in a very complicated  matrix\footnote{$l,m$  label the rows and $l',m'$ the columns.}$\langle a_{lm} a^*_{l'm'}\rangle$.

In this paper we explore the form of the two point correlation $\langle \delta( {\bf k})\delta ({\bf q})\rangle$ if translational invariance is broken by the presence of   cosmic string that passes through our horizon volume during inflation. We will assume that the string becomes unstable and disappears near the end of inflation and approximate the string as infinitely long and having infinitesimal thickness. In that case rotational invariance about the string axis and translational invariance along the string direction are left unbroken. Aligning the preferred direction  with the $ z$ axis these symmetries imply that the two point correlation of energy density correlations takes the form,
\begin{align}
\label{stringgeneral}
\langle \delta( {\bf k})\delta ({\bf q})\rangle= &(2 \pi)\delta(k_z+q_z)e^{i({\bf k}_{\perp}+{\bf q}_{\perp})\cdot {\bf x}_0}\nn
&P(k_{\perp},q_{\perp}, k_z, {\bf k}_{\perp} \cdot {\bf q}_{\perp}),
\end{align}
with $P$ symmetric under interchange of ${\bf k}_{\perp}$ and ${\bf q}_{\perp}$.  Here we have decomposed the wave vectors along the $z$ axis and the two dimensional subspace perpendicular to that is denoted by a subscript $\perp$.  ${\bf x}_0$ is a point on the string.  If the preferred direction is along an arbitrary direction $\hat {\bf n}=R \hat{\bf z}$, where $R$ is a rotation that leaves the point ${\bf x}_0$ fixed, then on the right hand side of Eq.~(\ref{stringgeneral}) the wave vectors are replaced by the rotated ones;  ${\bf k} \rightarrow R{\bf k}$ and ${\bf q} \rightarrow R{\bf q}$. The goal of this paper is to derive an explicit expression for the function $P(k_{\perp},q_{\perp}, k_z, {\bf k}_{\perp} \cdot {\bf q}_{\perp})$.

Using cylindrical spatial coordinates the metric for the inflationary spacetime with an infinitely long infinitesimally thin straight string directed along $z$ direction and passing through the origin is \cite{Abbassi}
\begin{equation}
  {\rm d}s^2=- {\rm d}t^2+a(t)^2\left[  {\rm d}\rho^2+\rho^2(1-4G\mu)^2 {\rm d}\theta^2+ {\rm d}z^2\right],
\end{equation}
where $a(t)=e^{Ht}$ is just the ordinary inflationary scale factor and $\mu$ is the tension along the string. We  compute the Fourier transform of the two point correlation of $\chi$ in this space-time. This is a simplified model for inflation where $\chi$ plays the role of the inflaton and $\delta ({\bf k}) \propto \chi ({\bf k})$. We focus on the cosmic string case because of the simplicity of the metric and not because of a strong physical motivation. Unless there is `` just enough  inflation" it is very unlikely that there would be a cosmic string in our horizon volume during inflation. If there was just enough inflation
\cite{Deruelle,{Goldwirth:1,{Goldwirth:2}}}
there could be other sources of violations of translational and rotational invariance
\cite{Aguirre:2007an,{Chang:2008gj},{Aguirre:2007wm},{Aguirre:2008wy},{Chang:2007eq},{Freivogel:2009it}}.
However, the cosmic string case does provides a simple physical model where the form of the violation of translational and rotational invariance can be explicitly calculated and it depends on only the parameter $G\mu$ and three other parameters that specify the location and alignment of the string. Real cosmic strings have a  thickness  of order $1/\sqrt{\mu}$ and so for it to be treated as thin we need $H^2<<\mu$ which implies that the dimensionless quantity $\epsilon =G\mu$ is much greater than, $G H^2$.

It is also possible to violate translational invariance by a point defect that existed during the inflationary era. In the conclusions we briefly  discuss  how the cosmic string case differs from the case of  a black hole located in our horizon volume during the inflationary era \cite{HTC}.

\widetext
\section{The Two Point Correlation Function of a Massless Scalar}
The metric for an inflationary spacetime with an infinitely long string along $z$ direction and through the origin is taken of the form \cite{Abbassi}
\begin{equation}
 {\rm d}s^2=- {\rm d}t^2+a(t)^2\left[  {\rm d}\rho^2+\rho^2(1-4G\mu)^2 {\rm d}\theta^2+ {\rm d}z^2\right],
\end{equation}
where $a(t)=e^{Ht}$ is just the ordinary inflationary scale factor. We let $\alpha=1-4G\mu$. In these coordinates the Lagrangian density for a massless scalar field is
\begin{eqnarray}
 \mathcal{L}_\chi&=&-\frac{\sqrt{-g}}{2}g^{\mu\nu}\partial_\mu\chi\partial_\nu\chi \\
                 &=&\frac{a(t)^3}{2}\rho\alpha\pbra{\frac{\partial\chi}{\partial t}}^2-\frac{a(t)}{2}\rho\alpha\pbra{\frac{\partial\chi}{\partial z}}^2-\frac{a(t)}{2}\rho\alpha\pbra{\frac{\partial\chi}{\partial\rho}}^2-\frac{a(t)}{2\rho\alpha}\pbra{\frac{\partial\chi}{\partial \theta}}^2.
\end{eqnarray}
The Hamiltonian\footnote{The same symbol is used for the Hamiltonian and Hubble constant during inflation, however the meaning of the symbol should be clear from the context.} is,
\begin{eqnarray}
 H&=&\int  {\rm d}^3 x(\pi\dot\chi-\mathcal{L})\\
            &=&\int ~\rho  {\rm d}\rho  {\rm d}\theta  {\rm d}z~ \frac{\alpha}{2}\left[ a(t)^3\left({\partial \chi \over \partial t}\right)^2+a(t)\pbra{\frac{\partial\chi}{\partial z}}^2+a(t)\pbra{\frac{\partial \chi}{\partial\rho}}^2+\frac{a(t)}{\rho^2\alpha^2}\pbra{\frac{\partial\chi}{\partial\theta}}^2\right].
\end{eqnarray}
It is convenient to introduce the  conformal time,
\begin{equation}
\tau=-{1 \over H} e^{-H t}
\end{equation}
and as $t$ goes from $-\infty$ to $\infty$ the conformal time $\tau$ goes from $-\infty$ to $0$. Since the metric only differs from de Sitter space by the presence of a conical singularity at $\rho=0$ the (equal time) two-point correlation of $\chi$ can easily shown to be,
\begin{equation}
\langle\chi(\rho,\theta,z,\tau) \chi(\rho',\theta',z',\tau) \rangle=\int_0^\infty{ {\rm d} k_{\perp}\over 2 \pi} k_{\perp}\int_{-\infty}^{\infty}{{\rm d}k_z \over 2\pi} e^{ik_z(z-z')}\sum_{m=-\infty}^{\infty} {e^{im(\theta-\theta')} \over 2 \pi}J_{|m/\alpha|}(k_{\perp}\rho)J_{|m/\alpha|}(k_{\perp}\rho'){|\chi_k(\tau)|^2 \over \alpha}.
\end{equation}
Here $\chi_k(\tau)$ are the usual mode functions in de Sitter space,
\begin{equation}
\label{mode}
\chi_k(\tau)=\frac{H}{\sqrt{2k}}e^{-ik\tau}\pbra{\tau-\frac{i}{k}}.
\end{equation}
We are interested in the late time, $k\tau \rightarrow 0$ behavior. Using the explicit form of $\chi_k(\tau)$ above,
\begin{equation}
\label{exact} \langle\chi(\rho,\theta,z,0) \chi(\rho',\theta',z',0)
\rangle={H^2 \over 2 \alpha}\int_0^\infty{ {\rm d} k_{\perp}\over 2
\pi} k_{\perp}\int_{-\infty}^{\infty}{{\rm d}k_z \over 2\pi}
e^{ik_z(z-z')}\sum_{m=-\infty}^{\infty} {e^{im(\theta-\theta')}
\over 2
\pi}{J_{|m/\alpha|}(k_{\perp}\rho)J_{|m/\alpha|}(k_{\perp}\rho')\over(k_{\perp}^2+k_z^2)^{3/2}}.
\end{equation}
The observed universe is consistent with the standard predictions of the inflationary cosmology. Therefore the violation of translational invariance due to the string is  a small perturbation,  and is  parametrized by the small quantity $\epsilon=4G\mu$. There are two approaches to calculate the Fourier transform of the two point correlation of $\chi$. One is to  expand the Bessel functions in Eq.~(\ref{exact}) about $\epsilon=0$ and then change to cartesian coordinates. Another approach, which is the one we take, is to abandon the exact result in Eq.~(\ref{exact}) and just do quantum mechanical perturbation theory about the unperturbed, $\epsilon=0$ background, {\it i.e.}, de Sitter space.

We need to compute the two-point correlation function $\left\langle \chi(\mathbf{x},t)\chi(\mathbf{y},t)\right\rangle$. Treating $\epsilon$ as a small perturbation, to first order of $\epsilon$, (see Ref.~\cite{Weinberg})
\begin{equation}
\label{perturbation}
 \left\langle \chi(\mathbf{x},t)\chi(\mathbf{y},t)\right\rangle\simeq \left\langle \chi_I(\mathbf{x},t)\chi_I(\mathbf{y},t)\right\rangle+i\int_{-\infty}^t~{ \rm d}t'e^{-\epsilon ' |t'|}\left\langle \left[ {H}_I(t'),\chi_I(\mathbf{x},t)\chi_I(\mathbf{y},t)\right] \right\rangle ,
\end{equation}
where $\epsilon '$ is an infinitesimal parameter that cuts off the early time part of the integration. In this case the interaction-picture Hamiltonian ${H}_I(t)$ is given by
\begin{eqnarray}
{H}_I&=&\int~\rho {\rm d}\rho {\rm d}\theta {\rm d}z \pbra{-\frac{\epsilon}{2}}\left[ a^3\left({\partial \chi_I \over \partial t}\right)^2+a\pbra{\frac{\partial\chi_I}{\partial z}}^2+a\pbra{\frac{\partial\chi_I}{\partial \rho}}^2-\frac{a}{\rho^2}\pbra{\frac{\partial\chi_I}{\partial \theta}}^2\right] \\
              &=&-\epsilon{H}_0+\epsilon\int\rho {\rm d}\rho{\rm  d}\theta dz\frac{a}{\rho^2}\pbra{\frac{\partial \chi_I}{\partial\theta}}^2,
\end{eqnarray}
where the interaction picture field $\chi_I$ has its time evolution governed by the unperturbed Hamiltonian,
\begin{equation}
{H}_0=\int\rho {\rm d}\rho {\rm d}\theta {\rm d}z\frac12\left[ a^3\left({\partial \chi_I \over \partial t}\right)^2+a\pbra{\frac{\partial\chi_I}{\partial z}}^2+a\pbra{\frac{\partial\chi_I}{\partial \rho}}^2+\frac{a}{\rho^2}\pbra{\frac{\partial\chi_I}{\partial \theta}}^2\right].
\end{equation}
Because we are interested in the effects that violate rotational and/or translational invariance in, $\Delta \left\langle \chi(\mathbf{x},t)\chi(\mathbf{y},t)\right\rangle=\left\langle \chi(\mathbf{x},t)\chi(\mathbf{y},t)\right\rangle-\left\langle \chi_I(\mathbf{x},t)\chi_I(\mathbf{y},t)\right\rangle$, we will drop the term proportional to $H_0$ in the interaction Hamiltonian leaving us with,
\begin{equation}
 {H}_I=\epsilon\int\rho {\rm d}\rho {\rm d}\theta {\rm d}z\frac{a}{\rho^2}\pbra{\frac{\partial \chi_I}{\partial\theta}}^2,
\end{equation}
to first order in $\epsilon$. The free field obeys the unperturbed equation of motion,
\begin{equation}
\frac{d^2\chi_I}{{\rm d}t^2}+3H\frac{{\rm d}\chi_I}{{\rm d}t}-\frac{1}{a(t)^2}\frac{{\rm d}^2\chi_I}{{\rm d}\mathbf{x}^2}=0.
\end{equation}
Upon quantization, $\chi_I$ becomes a quantum operator
\begin{eqnarray}
\chi_I(\mathbf{x},\tau)&=&\int\frac{d^3k}{(2\pi)^3}e^{i\mathbf{k}\cdot\mathbf{x}}\bbra{\chi_k(\tau)\beta(\mathbf{k})+\chi_k^*(\tau)\beta^\dag(-\mathbf{k})}\\
                       &=&\int\frac{d^3k}{(2\pi)^3}e^{ik_zz}e^{ik_\bot\rho\cos(\theta-\theta_k)}\bbra{\chi_k(\tau)\beta(\mathbf{k})+\chi_k^*(\tau)\beta^\dag(-\mathbf{k})}
\end{eqnarray}
where  $\chi_k(\tau)$ is given by Eq.~(\ref{mode}).
Note that we have converted to the conformal time $\tau=-e^{-Ht}/H$ and used cylindrical coordinates for ${\bf k}$ and ${\bf x}$ in the exponential. $\beta(\mathbf{k})$ annihilates the vacuum state and  satisfies the usual commutation relations, $[\beta({\bf k}), \beta^{\dagger}({\bf q})]=(2 \pi)^3\delta({\bf k} -{\bf q})$.
Combining these definitions that interaction Hamiltonian can be written in terms of creation and annihilation operators as,
\begin{eqnarray}
 {H}_I(\tau')=\epsilon\pbra{\frac{1}{H\tau'}}\lefteqn{\int\frac{d^3k}{(2\pi)^3}\int\frac{d^3q}{(2\pi)^3}\int d^3x'~e^{i\mathbf{k}\cdot\mathbf{x'}+i\mathbf{q}\cdot\mathbf{x'}}\frac{(y'k_x-x'k_y)(y'q_x-x'q_y)}{x'^2+y'^2}}\nn
&& \bbra{\chi_k(\tau')\beta(\mathbf{k})+\chi_k^*(\tau')\beta^\dag(-\mathbf{k})}\bbra{\chi_q(\tau')\beta(\mathbf{q})+\chi_q^*(\tau')\beta^\dag(-\mathbf{q})}.
\end{eqnarray}

Next we use the above results to compute the needed commutator,  $\vev{\left[{H}_I(\tau'),\chi_I(\mathbf{x},\tau)\chi_I(\mathbf{y},\tau)\right]}=\vev{\left[ {H}_I(\tau'),\chi_I(\mathbf{x},\tau)\right]\chi_I(\mathbf{y},\tau)}+\vev{\chi_I(\mathbf{x},\tau)\left[ {H}_I(\tau'),\chi_I(\mathbf{y},\tau)\right]}$. This gives,
\begin{eqnarray}
\label{interaction}
\vev{\left[ {H}_I(\tau'),\chi_I(\mathbf{x},\tau)\chi_I(\mathbf{y},\tau)\right]}&=&\frac{H^3\epsilon}{2\tau'}\int\frac{d^3k}{(2\pi)^3}\int\frac{d^3q}{(2\pi)^3}\int d^3x'~e^{i\mathbf{k}\cdot(\mathbf{x'}-\mathbf{x})+i\mathbf{q}\cdot(\mathbf{x'}-\mathbf{y})} {(y'k_x-x'k_y)(y'q_x-x'q_y) \over (x'^2+y'^2)}\nonumber  \\
&&{ 1 \over kq}\left[e^{-i(k+q)(\tau'-\tau)}\pbra{\tau'-\frac{i}{k}}\pbra{\tau'-\frac{i}{q}}\pbra{\tau+\frac{i}{k}}\pbra{\tau+\frac{i}{q}}-{\rm h}.{\rm c}.\right].
\end{eqnarray}
Converting the integration over $t'$ in Eq.~\eqref{perturbation} to the integration over the conformal time $\tau'$ $\pbra{{\rm d}t'=-\dfrac{{\rm d} \tau'}{H\tau'}}$, using Eq.~\eqref{interaction}, and noting that cutoff involving $\epsilon '$  removes the influence at the very early time, we  integrate over $\tau'$ to get
\begin{eqnarray}
\Delta\langle \chi(\mathbf{x},\tau)\chi(\mathbf{y},\tau)\rangle&=& -H^2\epsilon\int\frac{d^3k}{(2\pi)^3}\int\frac{d^3q}{(2\pi)^3}{\int d^3x'~e^{i\mathbf{k}\cdot(\mathbf{x'}-\mathbf{x})+i\mathbf{q}\cdot(\mathbf{x'}-\mathbf{y})}} \nonumber \\
&&  {(y'k_x-x'k_y)(y'q_x-x'q_y) \over x'^2+y'^2}
 \bbra{\frac{kq+q^2+k^2+k^2q^2\tau^2}{k^3q^3(k+q)}}.
\end{eqnarray}
Using,
\begin{equation}
\frac{(y'k_x-x'k_y)(y'q_x-x'q_y)}{x'^2+y'^2}=\mathbf{k}_\bot\cdot\mathbf{q}_\bot-\frac{(\mathbf{x'_\bot}\cdot\mathbf{k}_\bot)(\mathbf{x'_\bot}\cdot\mathbf{q}_\bot)}{{x'_\bot}^2},
\end{equation}
 gives
\begin{eqnarray}
\Delta\left\langle \chi(\mathbf{x},\tau)\chi(\mathbf{y},\tau)\right\rangle=-H^2\epsilon\lefteqn{\int\frac{d^3k}{(2\pi)^3}\int\frac{d^3q}{(2\pi)^3}e^{-i\mathbf{k}\cdot\mathbf{x}-i\mathbf{q}\cdot\mathbf{y}}2\pi\delta(k_z+q_z)\bbra{\frac{kq+q^2+k^2+k^2q^2\tau^2}{k^3q^3(k+q)}}}\nn
&&\bbra{\int d^2x'_\bot~e^{i(\mathbf{k_\bot}+\mathbf{q_\bot})\cdot\mathbf{x'_\bot}}\pbra{\mathbf{k_\bot}\cdot\mathbf{q_\bot}-\frac{(\mathbf{x'_\bot}\cdot\mathbf{k}_\bot)(\mathbf{x'_\bot}\cdot\mathbf{q}_\bot)}{{x'_\bot}^2}}},
\end{eqnarray}
where $x_{\perp}'=|{\bf x_{\perp}'}|$. It remains to perform the integration over $x'$.  We find that,
\begin{equation}
\int {\rm d^2}x'_{\perp}e^{i{\bf p _{\perp}}\cdot {\bf x'_{\perp}}} {x'_{\perp i}x'_{\perp j} \over{\  x'_{\perp}}^2}=(2 \pi)^2\delta({\bf p_{\perp}}){\delta_{ij} \over 2}+{4 \pi \over  {  p_{\perp}}^2}\left({\delta_{ij} \over 2}-{p_{\perp i }p_{\perp j}\over { p_{\perp} }^2}\right),
\end{equation}
and so
\begin{eqnarray}
\label{bigone}
\Delta \langle \chi(\mathbf{x},\tau)\chi(\mathbf{y},\tau)\rangle=-H^2\epsilon{\int\frac{d^3k}{(2\pi)^3}\int\frac{d^3q}{(2\pi)^3}e^{-i\mathbf{k}\cdot\mathbf{x}-i\mathbf{q}\cdot\mathbf{y}}2\pi\delta(k_z+q_z){\frac{kq+q^2+k^2+k^2q^2\tau^2}{k^3q^3(k+q)}}}   \nonumber   \\
\left[{{\bf k_{\perp}} \cdot {\bf q_{\perp}} \over 2}(2 \pi)^2\delta({\bf k_{\perp}} +{\bf q_{\perp}} )-{4 \pi \over ({\bf k_{\perp}}+{\bf q_{\perp}})^2} \left( {{\bf k_{\perp}} \cdot {\bf q_{\perp}} \over 2}  -{ {\bf k_{\perp}}\cdot ({\bf k_{\perp}}+{\bf q_{\perp}}){\bf q_{\perp}}\cdot ({\bf k_{\perp}}+{\bf q_{\perp}}) \over ({\bf k_{\perp}}+{\bf q_{\perp}})^2}\right)           \right].
\end{eqnarray}
The second term in the large square brackets of Eq.~(\ref{bigone}) appears naively to give rise to a logarithmic divergence in the integrations over $q$ and $k$ near ${\bf p_{\perp}}={\bf k_{\perp}}+{\bf q_{\perp}}=0$. However after doing the angular integration over the direction of ${\bf p_{\perp}}$ this potentially divergent term vanishes.

Writing the density perturbations as $\delta=\kappa \chi$  we arrive at  the following expression for the part of $P(k_{\perp},q_{\perp}, k_z, {\bf k}_{\perp} \cdot {\bf q}_{\perp})$ in Eq.~(\ref{stringgeneral}) that violates rotational and/or translational invariance,
\begin{align}
\label{yikes}
\Delta P(k_{\perp},q_{\perp}, k_z, {\bf k}_{\perp} \cdot {\bf q}_{\perp})=-\kappa^2 H^2 \epsilon\left({kq+q^2+k^2 \over k^3q^3(k+q)}\right) & \left[{{\bf k_{\perp}} \cdot {\bf q_{\perp}} \over 2}(2 \pi)^2\delta({\bf k_{\perp}} +{\bf q_{\perp}} )-{4 \pi \over ({\bf k_{\perp}}+{\bf q_{\perp}})^2} \left( {{\bf k_{\perp}} \cdot {\bf q_{\perp}} \over 2} \right.\right. \nonumber \\
&\left. \left. -{ {\bf k_{\perp}}\cdot ({\bf k_{\perp}}+{\bf q_{\perp}}){\bf q_{\perp}}\cdot ({\bf k_{\perp}}+{\bf q_{\perp}}) \over ({\bf k_{\perp}}+{\bf q_{\perp}})^2}\right)   \right].
\end{align}
In Eq.~(\ref{yikes})  $k=\sqrt{k_z^2+ k_{\perp}^2}$ and $q=\sqrt{k_z^2+ q_{\perp}^2}$. Eq.~(\ref{yikes}) is the main result of this paper. The dependence on the wave-vectors is scale invariant. However, the scale invariance is broken by the dependence on ${\bf x}_0$ that arises when one considers a string that doesn't pass through the origin. The first term in the large square brackets of Eq.~(\ref{yikes}) violates rotational invariance but not translational invariance. It is consistent with the form proposed by Ackerman {\it et. al.} \cite{Ackerman}. The second term in the large square brackets of Eq.~(\ref{yikes}) violates translational invariance.

Recall that in the model we have adopted the unperturbed density perturbations have a power spectrum $P_{0}(k)=\kappa^2H^2/(2k^3)$ and so $\epsilon$ characterizes the overall strength of the violations of rotational and translational invariance.  For $ ({\bf k_{\perp}}+{\bf q_{\perp}})\cdot {\bf x}_0>>1$ the exponential dependence on this quantity in Eq.~(\ref{stringgeneral})  oscillates rapidly and this suppresses the impact of $\Delta P$ on observable quantities which depend on integrals of $\langle  \delta({\bf k}) \delta({\bf q})\rangle$  over the  components of ${\bf q}$ and ${\bf k}$.
\section{Conclusions}
We have computed (with some simplifying assumptions)  the impact that an infinitely long and infinitesimally thin straight string that exists during inflation and passes through our horizon volume has on the perturbations of the energy density of the universe. We have assumed that the string disappears towards the end of the inflationary era. It may be possible to remove some of these assumptions or provide dynamics that realizes them. However, even without that,  Eq.~(\ref{yikes}) provides a simple and physically motivated  functional form (after modifying it so the string can have an arbitrary location and orientation) for the part of the density perturbations two point correlation function  that violates translational and rotational invariance.  It can be compared with data on the large scale structure of the universe and the anisotropy of the microwave background radiation.

Computations analogous to those performed in this paper can be done for a point defect (located at the origin $x=0$)  in de Sitter space using the metric \cite{Schwarzchild-deSitter,{Schwarzchild-FRW}},
\begin{equation}
{\rm d}s^2=-{\left(1-{r_0 \over a(t) x} \right) ^2\over \left(1+{r_0 \over a(t) x} \right) ^2} {\rm d}t^2+a(t)^2\left(1+{r_0 \over a(t) x} \right) ^4({\rm d}x^2 +x^2{\rm d}\Omega_2^2),
\end{equation}
where $a(t)=e^{Ht}$. In this case, perturbing about de Sitter space, yields the following interaction Hamiltonian for a massless scalar field $\chi$,
\begin{equation}
 H_I=4 r_0\int {\rm d}^3x \left({ 1 \over x}\right) a(t)^2\left( {{\rm d} \chi_I \over {\rm d} t } \right)^2=4 r_0\int{\rm d}^3 x\left( {1 \over x}\right)\left( {{\rm d} \chi_I \over {\rm d} \tau } \right)^2.
\end{equation}
For the point mass case the perturbative calculation of $\langle \chi({\bf x},t)\chi({\bf y},t)\rangle$ using Eq.~(\ref{perturbation}) has greater sensitivity to earlier times $t'$. For example, the factor of $e^{-\epsilon ' |t'|}$ does not regulate the $t'$ integration. An exponential regulator in conformal time would work but then the Fourier transform of the part of this two point function that violates translation invariance vanishes as $q\tau$ and $k \tau$ go to zero. This case was considered in Ref.~\cite{HTC}.

\section*{Acknowledgments}
We thank Sean Carroll for discussions particularly during the early stages of this work. We are grateful for the Department of energy for partial support under grant DE-FG03-92ER40701.

\end{document}